\documentclass[8.5pt,twoside,twocolumn]{article}
\usepackage[super,sort&compress,comma]{natbib}
\makeatletter
\newcommand*{\citenst}[2][]{%
  \begingroup
  \let\NAT@mbox=\mbox
  \let\@cite\NAT@citenum
  \let\NAT@space\NAT@spacechar
  \let\NAT@super@kern\relax
  \renewcommand\NAT@open{[}%
  \renewcommand\NAT@close{]}%
  \cite[#1]{#2}%
  \endgroup
}
\makeatother
\usepackage{mhchem}
\usepackage{times,mathptmx}
\usepackage{sectsty}
\usepackage{balance}

\usepackage{graphicx} %eps figures can be used instead
\usepackage{lastpage}
\usepackage[format=plain,justification=raggedright,singlelinecheck=false,font=small,labelfont=bf,labelsep=space]{caption}
\usepackage{fancyhdr}
\begin{document}

\thispagestyle{plain}
\fancypagestyle{plain}{
%\fancyhead[L]{\includegraphics[height=8pt]{headers/LH}}
%\fancyhead[C]{\hspace{-1cm}\includegraphics[height=20pt]{headers/CH}}
%\fancyhead[R]{\includegraphics[height=10pt]{headers/RH}\vspace{-0.2cm}}
\renewcommand{\headrulewidth}{1pt}}
\renewcommand{\thefootnote}{\fnsymbol{footnote}}
\renewcommand\footnoterule{\vspace*{1pt}%
\hrule width 3.4in height 0.4pt \vspace*{5pt}}

\newcommand{\um}{~\mu\mathrm{m}}
\setcounter{secnumdepth}{5}

\makeatletter
\def\subsubsection{\@startsection{subsubsection}{3}{10pt}{-1.25ex plus -1ex minus -.1ex}{0ex plus 0ex}{\normalsize\bf}}
\def\paragraph{\@startsection{paragraph}{4}{10pt}{-1.25ex plus -1ex minus -.1ex}{0ex plus 0ex}{\normalsize\textit}}
\renewcommand\@biblabel[1]{#1}
\renewcommand\@makefntext[1]%
{\noindent\makebox[0pt][r]{\@thefnmark\,}#1}
\makeatother
\renewcommand{\figurename}{\small{Fig.}~}
\sectionfont{\large}
\subsectionfont{\normalsize}

\fancyfoot{}
%\fancyfoot[LO,RE]{\vspace{-7pt}\includegraphics[height=9pt]{headers/LF}}
%\fancyfoot[CO]{\vspace{-7.2pt}\hspace{12.2cm}\includegraphics{headers/RF}}
%\fancyfoot[CE]{\vspace{-7.5pt}\hspace{-13.5cm}\includegraphics{headers/RF}}
%\fancyfoot[RO]{\footnotesize{\sffamily{1--\pageref{LastPage} ~\textbar  \hspace{2pt}\thepage}}}
%\fancyfoot[LE]{\footnotesize{\sffamily{\thepage~\textbar\hspace{3.45cm} 1--\pageref{LastPage}}}}
\fancyhead{}
\renewcommand{\headrulewidth}{1pt}
\renewcommand{\footrulewidth}{1pt}
\setlength{\arrayrulewidth}{1pt}
\setlength{\columnsep}{6.5mm}
\setlength\bibsep{1pt}

\twocolumn[
  \begin{@twocolumnfalse}
\noindent\LARGE{\textbf{A microscopic view of the yielding transition in concentrated emulsions}}
\vspace{0.6cm}

\noindent\large{\textbf{E. D. Knowlton,\textit{$^{a}$} D. J. Pine,$^{\ast}$\textit{$^{b}$} and L. Cipelletti,$^{\ast}$\textit{$^{c,d}$  }
}}\vspace{0.5cm}
%Please note that \ast indicates the corresponding author(s) but no footnote text is required.

\noindent\textit{\small{\textbf{Received Xth XXXXXXXXXX 20XX, Accepted Xth XXXXXXXXX 20XX\newline
First published on the web Xth XXXXXXXXXX 20XX}}}

\noindent \textbf{\small{DOI: 10.1039/b000000x}}
\vspace{0.6cm}
%Please do not change this text.

\noindent \normalsize{We use a custom shear cell coupled to an optical microscope to investigate at the particle level the yielding transition in concentrated emulsions subjected to an oscillatory shear deformation. By performing experiments lasting thousands of cycles on samples at several volume fractions and for a variety of applied strain amplitudes, we obtain a comprehensive, microscopic picture of the yielding transition. We find that irreversible particle motion sharply increases beyond a volume-fraction dependent critical strain, which is found to be in close agreement with the strain beyond which the stress-strain relation probed in rheology experiments significantly departs from linearity. The shear-induced dynamics are very heterogenous: quiescent particles coexist with two distinct populations of mobile and `supermobile' particles. Dynamic activity exhibits spatial and temporal correlations, with rearrangements events organized in bursts of motion affecting localized regions of the sample. Analogies with other sheared soft materials and with recent work on the transition to irreversibility in sheared complex fluids are briefly discussed.}
\vspace{0.5cm}
\end{@twocolumnfalse}
]

%Footnotes
%\footnotetext{\dag~Electronic Supplementary Information (ESI) available: [details of any supplementary information available should be included here]. See DOI: 10.1039/b000000x/}

%Please use \dag to cite the ESI in the main text of the article.
%If you article does not have ESI please remove the the \dag symbol from the title and the above footnotetext.

\footnotetext{\textit{$^{a}$~Department of Chemical Engineering, University of California Santa Barbara, Santa Barbara, CA 93106, USA}}
\footnotetext{\textit{$^{b}$~Center for Soft Matter Research, Department of Physics, New York University, New York, NY 10003, USA. E-mail: pine@nyu.edu}}
\footnotetext{\textit{$^{c}$~Department of Chemical and Biomolecular Engineering, Polytechnic School of Engineering, New York University, New York, NY 11201, USA. E-mail: pine@nyu.edu.}}
\footnotetext{\textit{$^{d}$~Universit\'{e} Montpellier 2, Laboratoire Charles Coulomb UMR 5221, F-34095, Montpellier,
France. E-mail: luca.cipelletti@univ-montp2.fr.}}
\footnotetext{\textit{$^{e}$~CNRS, Laboratoire Charles Coulomb UMR 5221, F-34095, Montpellier, France.}}

%additional addresses can be cited as above using the lower-case letters, c, d, e... If all authors are from the same address, no letter is required

\section{Introduction}
\label{sec:introduction}

Concentrated emulsions are viscoelastic materials, whose response to a mechanical perturbation depends both on the time scale and the amplitude of the perturbation.\cite{mason1996,mason1997a} For a small applied stress or strain, concentrated emulsions behave as a solid on short time scales, but eventually flow as liquids on much longer time scales. Similarly, a transition from a predominantly solid-like to a mostly liquid-like behavior is observed when varying the amplitude of an applied strain, for example. An elastic response is observed up to a critical strain, typically a few percent, beyond which the emulsion yields and flows as a fluid. This yielding transition has an obvious importance for the countless everyday materials based on emulsions, \textit{e.g.} in the food and personal care industry.

At a more fundamental level, the yielding transition of emulsions and other amorphous soft materials (concentrated colloids, foams, clays, pastes) has been extensively investigated as a prototypical manifestation of the (un)jamming transition~\cite{LiuNature1998} and to understand plasticity in amorphous solids. One of the key question concerns the relationship between the rheological behavior and the microscopic rearrangements that are ultimately responsible for the macroscopic mechanical behavior. While in crystalline solids it has been firmly established that yielding is mediated by defects,\cite{chen1992,biswas2013} identifying the microscopic elementary events leading to flow in amorphous materials is more challenging, due to the their disordered structure.

Theoretical work has introduced the notion that localized, non-affine rearrangements are the building blocks of the transition from elastic deformation to plastic flow, for example, in the free volume approach of the flow defects of Ref.~\citenst{spaepen1977}, or in the shear transformation zones of Ref.~\citenst{falk1998}. Subsequent simulations\cite{onuki1997,miyazaki2004,LemaitrePRL2009,boquet2009,TsamadosEPJE2010,BarratPRL2011} and experiments~\cite{besseling07,schall07,goyon2008,jop2012} have confirmed that plastic events are localized, focusing predominantly on the stationary flow resulting from an applied constant strain rate.

The microscopic behavior of amorphous viscoelastic materials under an oscillatory shear has received less attention, in spite of the fact that cyclic perturbations are widely used in rheological~\cite{Larson} and fatigue tests in material science. H\'{e}braud and coworkers~\cite{hebraud1997} have probed the microscopic dynamics of concentrated emulsions under a sinusoidal shear deformation using diffusing wave spectroscopy~\cite{DWSGeneral} (DWS), a dynamic light scattering technique in the limit of strong multiple scattering. The main finding of this work was that plastic rearrangements are highly heterogeneous: the majority of the sample is deformed elastically and recovers its microscopic configuration after a full cycle, while a small fraction of drops undergo irreversible rearrangements. These rearrangements repeatedly affect the same subpopulation of drops. As the applied strain, $\gamma$, is increased, the fraction of rearranged sample increases. Quite surprisingly, at the yielding transition (as identified by rheology) more than 90\% of the sample still behaves elastically at the microscopic level. Petekidis \textit{et al.}~\cite{PetekidisPRE2002} have applied the same method to glassy suspensions of colloidal hard spheres. They also find heterogeneous plastic rearrangements, but the populations of mobile and immobile (after one full cycle) particles were found to evolve with time. As for the emulsions, the fraction of particles undergoing plastic events increases with $\gamma$. Interestingly, the sharpness of the yielding transition was found to depend on colloid volume fraction, $\varphi$: the higher $\varphi$, the more abrupt the transition.

Although these works pioneered the investigation of the yielding transition at a microscopic level, they suffered to some extent from the difficulties inherent to the interpretation of DWS data. The method provides information on the \textit{average} microscopic dynamics, but provides no direct information on the heterogeneous nature of the rearrangements. Various assumptions have to be made, which cannot be verified independently. For example, it is extremely difficult to discriminate between a partial loss of correlation of the scattered light due to substantial motion of a small fraction of the sample, or, conversely, stemming from more restrained displacements affecting a larger fraction of the sample. Furthermore, no information on the temporal organization of the rearrangements can be obtained, since the intensity correlation function measured in DWS usually needs to be averaged over time. This is an important limitation, since numerical simulations~\cite{Priezjev2013} have shown that rearrangements occur intermittingly in a three-dimensional Lennard-Jones amorphous solid, a model system for soft particles. In this respect, optically microscopy is an attractive alternative to DWS, and it has been recently used to probe the yielding transition under oscillatory shear in both colloids~\cite{ChenPRE2010} and emulsions~\cite{Clara-Rahola2012}. The authors of Ref.~\citenst{Clara-Rahola2012} focus mainly on the motion of drops that \textit{do} recover their initial position after a full shear cycle; however, an interesting side observation that supports the conclusions of~\cite{hebraud1997} is the fact that only a small fraction of the sample undergoes irreversible rearrangements, even close to the macroscopic yielding transition.

Oscillatory strain experiments have also been considered under the perspective of the transition from reversible to irreversible motion. In Ref.~\citenst{PetekidisPRE2002} it was already noted that cyclic deformations may be fully reversible because the hydrodynamics equations at low Reynolds number are invariant under time reversal, leading to a `reversible viscous distortion'. Experiments on relatively diluted, non-Brownian suspensions have explored systematically this effect~\cite{pine2005,corte2008}, showing that a transition exists between a low-$\gamma$ absorbing state where particles fully recover their initial position after one shear cycle, and an irreversible state, where particles undergo diffusive motion as the result of collisions with other colloids for large enough strain amplitudes. In Refs.~\citenst{pine2005,corte2008} the hallmarks of a second-order transition have been observed: the diffusivity increases gradually beyond a critical strain $\gamma_\mathrm{c}$ separating the reversible and irreversible regimes, and the time required to attain a stationary state seemingly diverges as  $\gamma$ approaches $\gamma_\mathrm{c}$ from both below and above. Remarkably, a similar reversible-irreversible transition has been reported recently for concentrated systems of interacting particles, both numerically~\cite{SastryPRE2013} (three-dimensional Lennard-Jones glass) and experimentally (grains~\cite{SlotterbackPRE2012} and repulsive colloids in two dimensions~\cite{keim2013}). This is particularly surprising given that in Refs.~\citenst{pine2005,corte2008} the reversible state is due to the absence of particle interactions at small $\gamma$, an ingredient that is clearly missing in Refs.~\citenst{SastryPRE2013,SlotterbackPRE2012,keim2013}. Quite intriguingly, the authors of a very recent experimental study on a dilute system with long-range interactions~\cite{jeanneret2014} reach an opposite conclusion: a reversible-irreversible transition is observed, but the transition is first-order. These contrasting results indicate that our understanding of sheared amorphous systems is still far from complete and calls for new investigations.

In this paper, we report on an optical microscopy study of the dynamics of concentrated emulsions submitted to an oscillatory shear. We present data for five volume fractions, from just above random close packing to $\varphi = 0.88$, where the drops are highly compressed. For each $\varphi$, we perform experiments at a variety of strain amplitudes, both below and above the yielding transition as inferred from rheology. Each experimental run consists of several thousands of shear cycles, thereby allowing a thorough investigation of the spatio-temporal behavior of plasticity. Finally, microscopy observation are systematically compared to the results of rheological tests, in order to establish a connection between the microscopic dynamics and the mechanical response across the yielding transition. The rest of the paper is organized as follows: in Sec.~\ref{sec:methods} we describe the emulsion preparation, the shear cell, and the image analysis procedure. Section.~\ref{sec:resdisc} presents our results, which are discussed and summarized in Sec.~\ref{sec:conclusions}.

\section{Materials and methods}
\label{sec:methods}

\subsection{Emulsion preparation}
We use oil in water emulsions, where the oil phase is polydimethilsyloxane, PDMS (Petrarch) and the dispersing phase a mixture of water and glycerol, to partially match the refractive index of the drops. The content of glycerol is fine-tuned by optimizing the optical contrast for microscopy observations. Since the optical properties of the emulsions depend on compression, different amounts of glycerol were added depending on the emulsion volume fraction. For $\varphi < 0.80$, the glycerol/water ratio is 40\% w/w; it is 20\% for $0.80 \le \varphi \le 0.85$, while for the most concentrated emulsion ($\varphi = 0.88$) the ratio is 10\%. The drops are stabilized by a non-ionic surfactant. Various surfactants have been tested in order to achieve good stability and avoid wall slip under shear. The best results are obtained with TMN-10 (2,6,8 trimethylnonyl ethoxide, Union Carbide), at a concentration of 1.7\% w/w in the aqueous phase.

The emulsions are prepared as described in Refs.~\citenst{mason1997,mabille2000}. The method consists of preparing a crude emulsion (`premix') by slowly dripping PDMS in the aqueous phase in the presence of an excess of surfactant, while gently stirring. The premix is then sheared at a high strain rate (typically, $1500~\mathrm{s}^{-1}$) in a couette cell, to reduce the drop size down to $2.4\um$, with a relative size polydispersity of 20\%, as measured by static light scattering (Malvern Mastersizer). Solvent is added to reduce the surfactant to the desired final value, and the emulsion is finally concentrated to the desired volume fraction by centrifugation.  The precise value of $\varphi$ is determined by weighting the fraction of oil left after evaporating the dispersing phase in a vacuum oven operated at $70~^\circ\mathrm{C}$.

\subsection{Microscopy under shear and image analysis}
\label{subsec:microscopy}

We use a custom-made shear cell and a Nikon Eclipse TE-2000 inverted microscope to visualize shear-induced rearrangements. The shear cell design is based on that by Petekidis \textit{et al.}~\cite{PetekidisPRE2002} In brief, the sample is confined in a gap of thickness $e = 100\um$ between two parallel borosilicate glass plates. The lower plate is a coverslip, the upper one a standard microscopy slide. To avoid wall slip, both plates are roughened using wet sandpaper and lapping compounds, resulting in a roughness of about $1-2\um$, as estimated by optical microscopy. A small window  of about $1~\mathrm{mm}^2$ is left un-roughened on the coverslip, to allow for visualization. A lever system allows the upper and lower plate to be moved in opposite directions, with a stagnation plane located approximately midway in the gap. A cyclic strain is imposed by a piezoelectric actuator (PiezoMechanik GmbH), whose position controller is driven by a sinusoidal signal (frequency $f=1$ Hz) issued from an ultra-low distortion function generator (Stanford Research Systems). The strain values quoted in the following are obtained from $\gamma = y_0/e$, with $y_0$ the amplitude of the relative motion of the two plates.

The shear cell is fixed on the stage of an inverted microscope equipped with differential interference contrast (DIC) optics and an oil-immersion 100x objective and oil-immersion condenser. The depth of field is $0.5\um$, much smaller than the drop size, and the imaged plane lays about $10\um$ from the coverslip. Images are taken with a Matrox Helios frame grabber and a Dalsa Pantera TF 1M60 CCD camera, triggered so as to take a strobed movie at one frame per shear cycle. The image size is 1024 x 1024 pixels and the pixel size corresponds to $0.12\um$ in the sample. Before any experiment at a given strain and volume fraction, the sample is presheared at a high strain, typically 250-300\% at a frequency of 1~Hz. The strain amplitude is then reduced to the target value and at least 300 cycles are imposed at the final $\gamma$ before starting the image acquisition.

\begin{figure}[h]
\centering
  \includegraphics[width=1\columnwidth,clip]{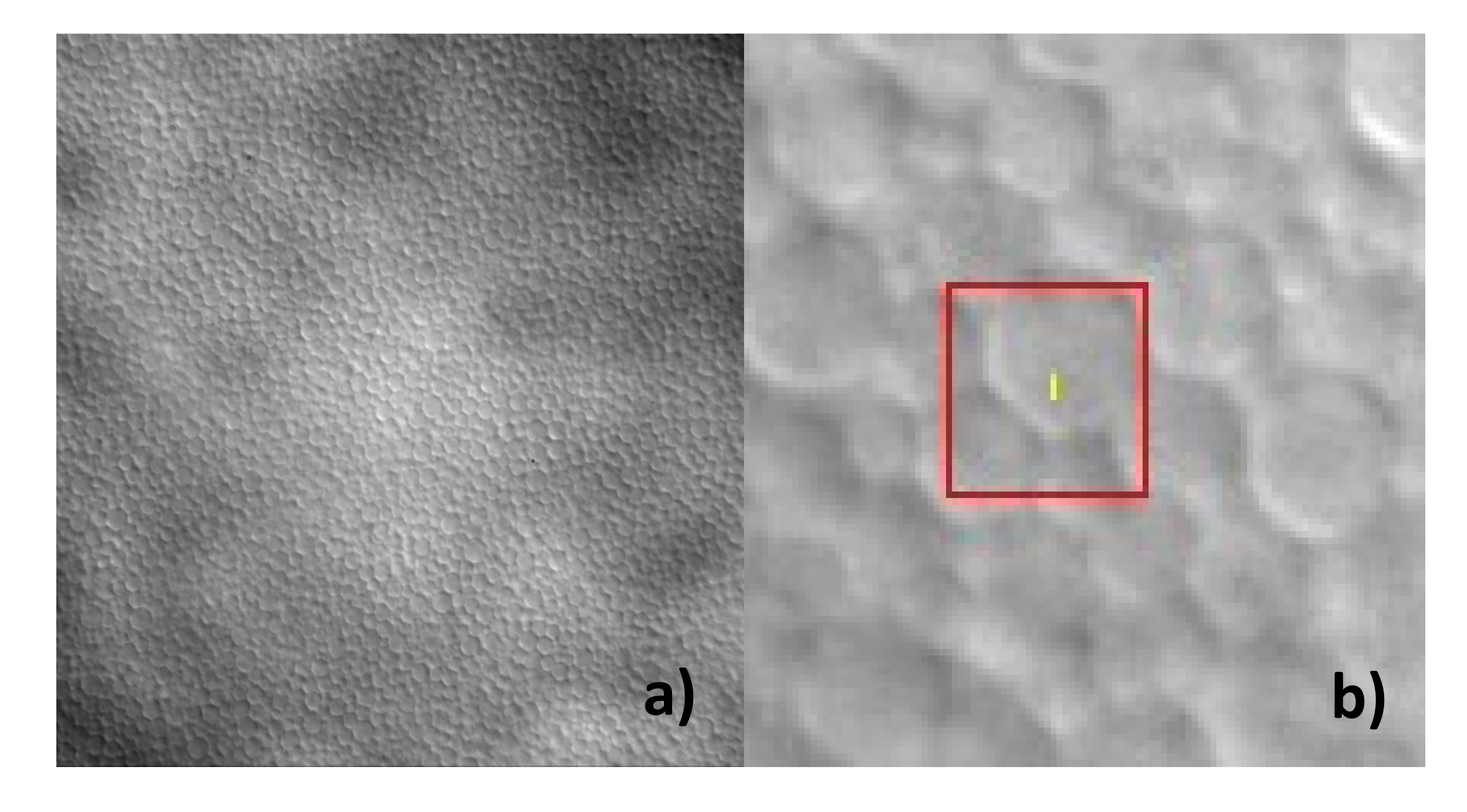}
  \caption{(a) DIC microscopy image of an emulsion at $\varphi = 0.65$. The field of view is $122.9\times122.9 \um^2$. The long-range intensity modulations are due to the illumination through a roughened microscopy slide. (b) zoom on a portion of the image shown in (a): the drops are clearly visible. The red square shows the size ($3.84 \times 3.84 \um^2$) of the ROIs used to measure the drop displacements by an image correlation algorithm. The yellow line corresponds to the $y$ component of the rms drop displacement after one shear cycle, for the largest applied shear amplitude, $\gamma = 7.88\%$.}
  \label{fig:drops}
\end{figure}

Figure~\ref{fig:drops}(a) shows a representative image of an emulsion at $\varphi = 0.65$, the lowest volume fraction that was investigated. Tracking the motion of individual drops is very difficult due to size polydispersity and because the drop shapes in compressed emulsions are not spherical. These difficulties could be attenuated by using larger drops, but this would limit the number of drops that can be visualized. We therefore
use a different approach and quantify motion using image correlation velocimetry (ICV),\cite{TokumaruExpInFluids1995} a cross-correlation method similar to particle imaging velocimetry.\cite{WesterweelMeasSciTech1997} Each image is divided in regions of interest (ROI) of size $32 \times 32~\mathrm{pixels}^2$; a displacement vector $\Delta\mathbf{ r}(\mathbf{R},t,\tau)$ is assigned to a ROI centered in $\mathbf{R}$ by maximizing the spatial cross-correlation between the intensity pattern of the ROI at time $t$ and that of an image taken at time $t+\tau$. More precisely, $\Delta\mathbf{r}$ is obtained with sub-pixel resolution by fitting the peak of $\mathrm{Corr}(\Delta;\mathbf{R},t,\tau)$ by a paraboloid, where
\begin{multline}
\mathrm{Corr}(\Delta;\mathbf{R},t,\tau) = \left < I(\mathbf{r},t)I(\mathbf{r}+\Delta,t+\tau)\right >_{\mathbf{R}} - \\
\left< I(\mathbf{r},t)\right >_{\mathbf{R}}\left< I(\mathbf{r}+\Delta ,t+\tau)\right >_{\mathbf{R}}\,,
\label{eq:crosscorr}
\end{multline}
where $\mathbf{r}$ is a pixel coordinate, $\Delta$ a spatial shift whose components are integer numbers of
pixels, and where $\left < \cdot \cdot \cdot \right >_{\mathbf{R}}$ indicates an average over pixels belonging to
the ROI centered in $\mathbf{R}$. In the following, time will be  expressed in numbers of shear cycles and only the dynamics over one cycle will be discussed, corresponding to $\tau = 1$ in Eq.~(\ref{eq:crosscorr}). The red box in Fig.~\ref{fig:drops}(b) shows the size of one of the ROIs used by the image correlation algorithm: it typically contains about three drops. As it will be discussed in the following, even at the largest applied strains the typical displacements over one shear cycle are relatively small, of order $0.2\um$. The yellow line in Fig.~\ref{fig:drops}(b) visualizes such a displacement in the $y$ direction. In principle, one may distinguish between motion along the direction parallel or perpendicular to the imposed strain ($x$ and $y$, respectively). We find no significant difference between the dynamics in the two directions, as also seen for colloids in Ref.~\citenst{ChenPRE2010}: in the following we discuss the results for the $y$ component of the displacement, which would be zero in the absence of plastic rearrangements, even if the images were not perfectly strobed with the shear cycle. By inspecting the results obtained for very low applied strains, where no irreversible rearrangements are expected to occur, we estimate the  error on the determination of the ROI displacement to be on the order of 0.05 pixel at most.

%------------------------------------------------------------------- FIG
%------------------------------------------------------------------ FIG

\section{Results}
\label{sec:resdisc}

\begin{figure}[h]
\centering
  \includegraphics[width=1\columnwidth,clip]{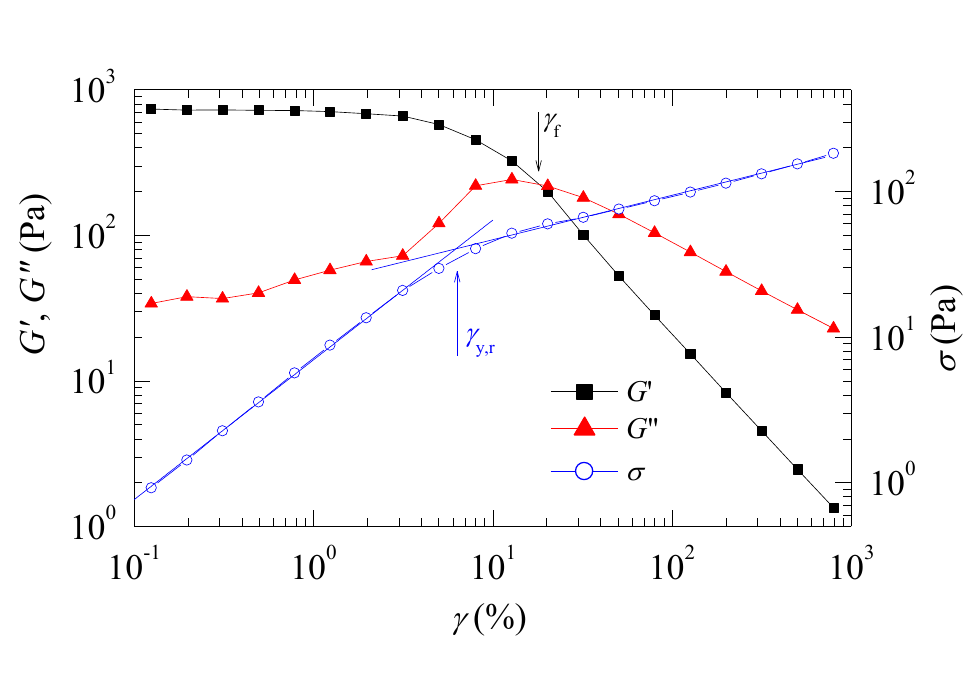}
  \caption{ Left axis and solid symbols: shear moduli \textit{vs.} strain amplitude $\gamma$, measured by oscillatory rheology at a frequency of 1 Hz, for an emulsion at volume fraction $\varphi = 0.83$. The cross-over or fluidization strain $\gamma_\mathrm{f}$ is defined by the intersection of $G'$ and $G''$ (black arrow). Right axis and open symbols: shear stress \textit{vs.} $\gamma$. The rheological yield strain $\gamma_\mathrm{y,r}$ is defined by the intersection of the two lines describing the behavior of $\sigma$ at small and large strain, respectively (blue arrow).}
  \label{fig:rheo}
\end{figure}

We start by discussing the macroscopic mechanical behavior of the emulsions~\cite{mason1996}, as measured by oscillatory rheology in a 50-mm cone-and-plate geometry. Figure~\ref{fig:rheo} shows the strain dependence of the elastic, $G'$, and loss, $G''$, moduli, measured at an oscillation frequency $f = 1$ Hz. Although we show here data for a single volume fraction, $\varphi = 0.83$, qualitatively similar behavior is observed for the whole range over which experiments were performed, $0.64 \le \varphi \le 0.88$. As the applied strain increases, $G''$ grows, indicating that dissipative processes, presumably due to drop rearrangements, become increasingly likely. However, the emulsions retain an overall solid-like behavior, with $G'$ larger than  $G''$ and only weakly dependent on $\gamma$. Beyond a few percent of applied strain amplitude, $G'$ drops significantly and $G''$ grows until $G' = G''$ at a `fluidization' strain $\gamma_\mathrm{f} \approx 20\%$. For $\gamma > \gamma_\mathrm{f}$, both moduli decrease with strain, but the loss modulus dominates, indicating fluid-like behavior. It should be emphasized that the transition from solid-like to fluid-like behavior is a very smooth one, extending over at least one decade in strain amplitude. Because of its smooth nature, several alternative ways are routinely used to define the yielding transition in concentrated emulsions. One criterion is based on the crossing point of the moduli, $\gamma_\mathrm{f}$, introduced above. A more stringent criterion is to identify the strain amplitude beyond which significant deviations from the linear regime are observed. To this end, we show in Fig.~\ref{fig:rheo} the $\gamma$ dependence of the shear stress $\sigma$ measured during the imposed strain sweep (open circles and right axis). Two power-law regimes are clearly visible, as evidenced by the straight lines. The crossover between the initial linear regime and the large-$\gamma$, sublinear growth of the stress defines a rheological yield strain,  $\gamma_\mathrm{y,r} = 6.4\%$, which occurs substantially earlier than the fluidization crossover.

%------------------------------------------------------------------- FIG
%------------------------------------------------------------------ FIG

\begin{figure}[h]
\centering
  \includegraphics[width=1\columnwidth,clip]{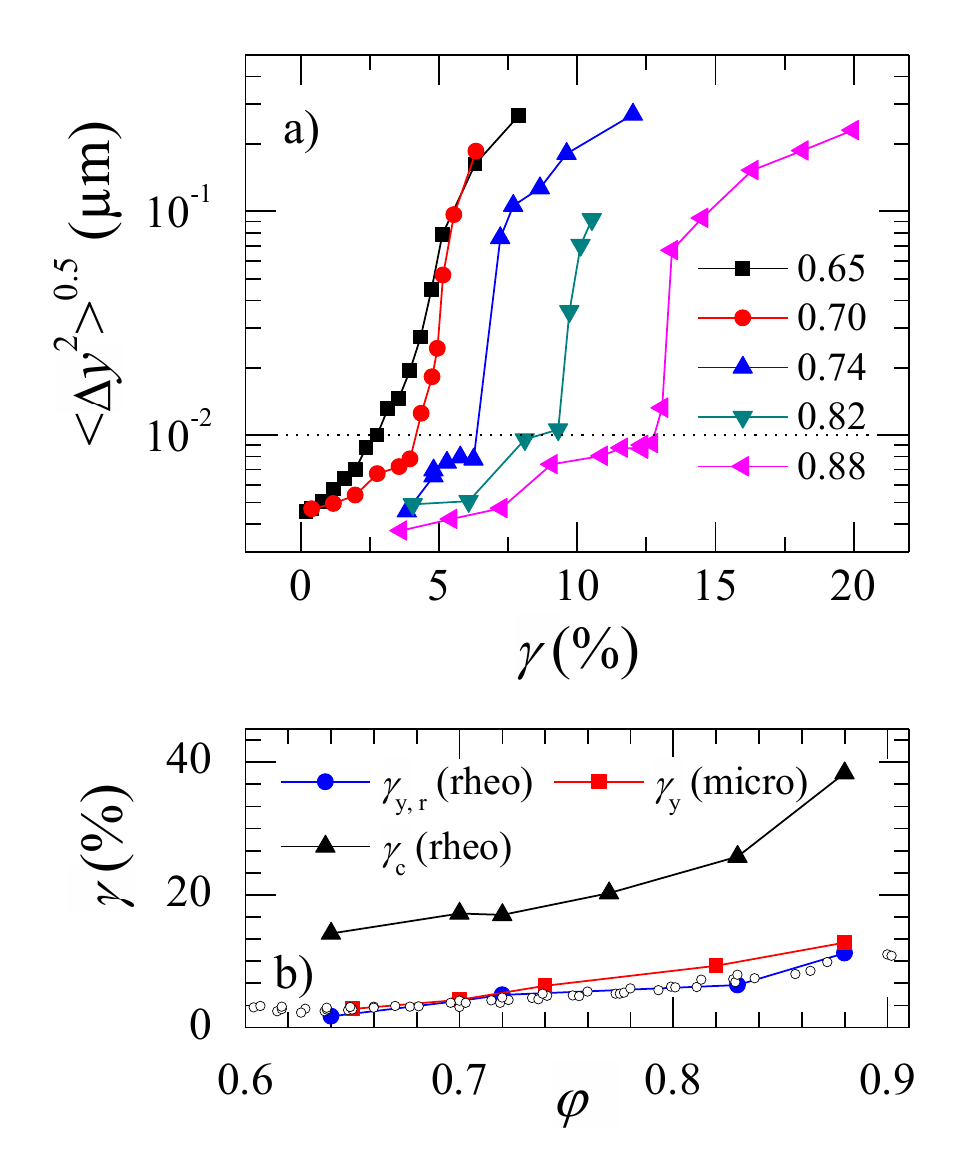}
  \caption{ (a) Shear dependence of the rms displacement per shear cycle in the $y$ direction, perpendicular to the applied shear. Data are labeled by the emulsion volume fraction. The dotted line is the threshold used to define the microscopic yield strain, $\gamma_\mathrm{y}$. (b) Volume fraction dependence of the yielding transition, as defined by various quantities. The small open symbols are rheology data taken from Ref.~\citenst{mason1996}.}
  \label{fig:transition}
\end{figure}

In order to investigate the yielding transition at a microscopic level, we measure the displacement field over one shear cycle, as a function of both emulsion volume fraction and strain amplitude. We show in Fig.~\ref{fig:transition}(a) $\sqrt{\langle\Delta y^2\rangle}$, the rms displacement in the $y$ direction, as a function of  $\gamma$ for five emulsions with volume fraction ranging from $\varphi = 0.65$, just above the jamming transition, up to $\varphi = 0.88$, where the emulsion is highly compressed. Data are averaged over all ROIs and over the duration of each experiment. A sharp increase of $\sqrt{\langle\Delta y^2
\rangle}$ is observed beyond a $\varphi$-dependent critical strain. This has to be contrasted to the much gentler transition observed by rheology (Fig.~\ref{fig:rheo}). For $\varphi \ge 0.74$, the transition is particularly abrupt. For example, at $\varphi = 0.74$ $\sqrt{\langle\Delta y^2\rangle}$ jumps by one order of magnitude when the strain increases by less than $1\%$. For the sake of comparison, we consider the variation of $\tan \delta = G''/G'$ around $\gamma_{\mathrm{y,r}}$, as $\tan \delta$ is often used to characterize the transition from solid-like to fluid-like behavior.\cite{Larson} For the system shown in Fig.~\ref{fig:rheo}, we find that $\tan \delta$ increases by a decade over a much wider interval of strain amplitude, $\Delta  \gamma \approx 16\%$. Interestingly, the yielding transition detected at a microscopic level becomes less sharp as the jamming transition is approached from above. This can be seen in Fig.~\ref{fig:transition}(a), where for the most diluted sample a one-decade growth of $\sqrt{\langle\Delta y^2\rangle}$ occurs over a strain interval of about $3\%$, three times larger than for samples at $\varphi \ge 0.74$. This suggests that the nature of the yielding transition may depend on the distance from the jamming transition.
%------------------------------------------------------------------- FIG
%------------------------------------------------------------------ FIG

We define the microscopic yield strain $\gamma_\mathrm{y}$ as the strain beyond which the rms displacement exceeds $10^{-2}\um$ (dotted line in Fig.~\ref{fig:transition}(a)). This threshold is chosen based on the behavior of the more concentrated samples, and is then used at all $\varphi$. The volume fraction dependence of $\gamma_y$ is shown in Fig.~\ref{fig:transition}(b), together with the two yield strains defined from rheological measurements, as discussed in reference to Fig.~\ref{fig:rheo}. Remarkably, we find that the yield strain inferred from microscopy measurements agrees very well with $\gamma_\mathrm{y,r}$ defined by the onset of non-linear behavior in the stress-strain relation. This result is particularly surprising given that the sharpness of the transition is very different for microscopy and rheology. The rheological yield strain $\gamma_\mathrm{y,r}$ has been shown to depend only marginally on sample details (drop size, composition, etc.), as shown by the open symbols in Fig.~\ref{fig:transition}(b), taken from a comprehensive set of experiments by Mason \textit{et al.}.\cite{mason1996} This suggests that the behavior observed in our microscopy experiments should be quite general. It should be noted that at all volume fractions $\gamma_\mathrm{y}$ is much smaller than the rheological cross-over strain $\gamma_\mathrm{c}$. Thus, a substantial increase in drop motion occurs well before the sample is actually fully fluidized.

%------------------------------------------------------------------- FIG
\begin{figure}[h]
\centering
  \includegraphics[width=1\columnwidth,clip]{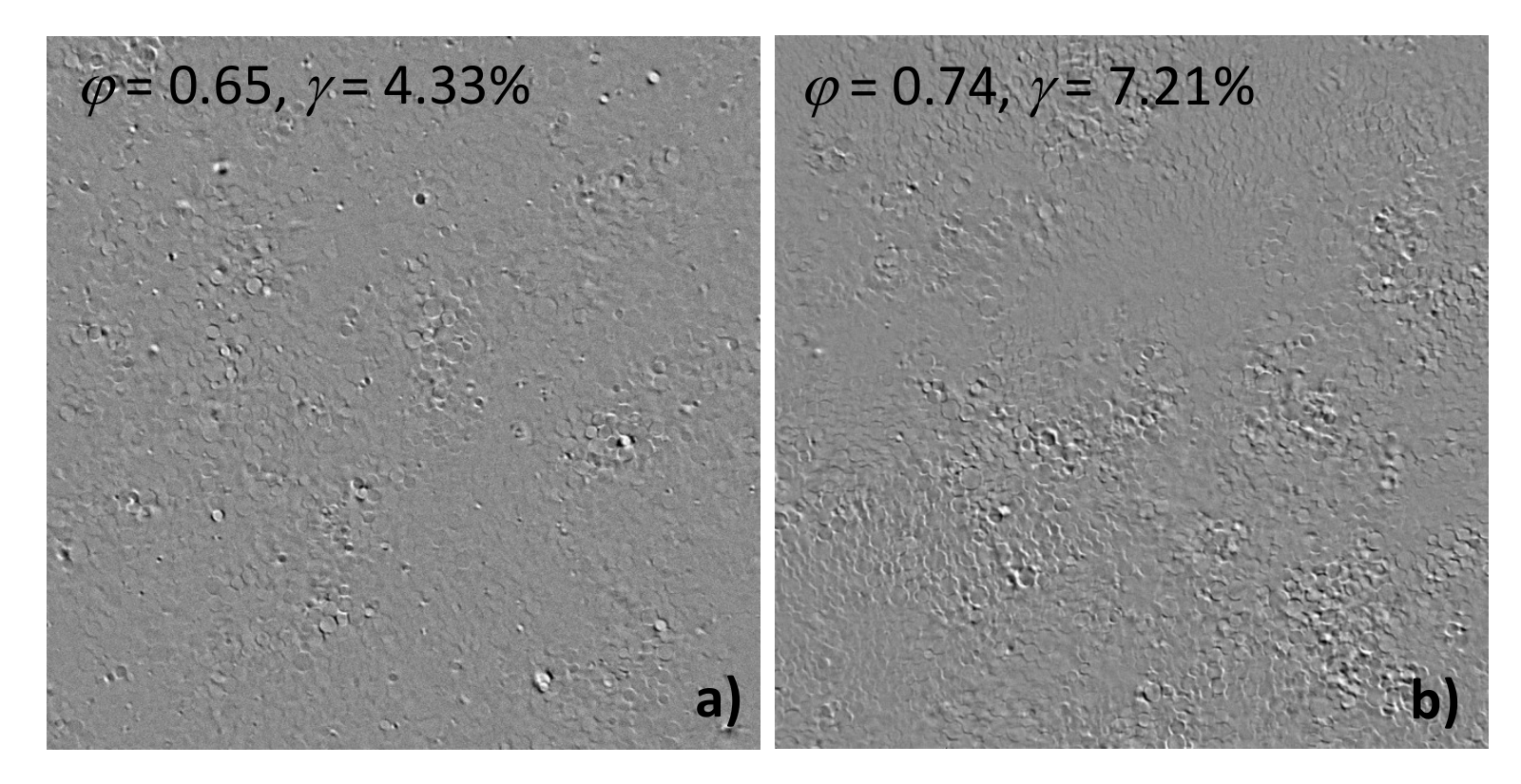}
  \caption{Difference between two successive strobed images: regions where motion occurred show up as features brighter or dimmer than the average background, while immobile regions are featureless. For each sample, the applied strain is just above the yield strain $\gamma_\mathrm{y}$.}
  \label{fig:differences}
\end{figure}
%------------------------------------------------------------------ FIG

While Fig.~\ref{fig:transition} quantifies the strain and $\varphi$ dependence of the average dynamics, our microscopy experiments allow us to investigate in detail both the spatial and temporal distribution of the rearrangements responsible for the drop displacement. A convenient way to visualize the regions that have undergone rearrangements over one cycle is to inspect the difference between a pair of successive strobed images. Figure~\ref{fig:differences} shows the result of such image subtraction for two representative volume fractions and for strain amplitudes just above $\gamma_\mathrm{y}$. Regions where no motion occurred appear identical in the pair of images, yielding patches of uniform grey level in Fig.~\ref{fig:differences}. By contrast, rearranged regions show up as contrasted zones, whose intensity changes from bright to dark along the contour of a displaced droplet. It is clear that the drop motion is spatially heterogenous, with islands of mobile particles separated by quiescent patches.

%------------------------------------------------------------------- FIG
\begin{figure}[h]
\centering
  \includegraphics[width=1\columnwidth,clip]{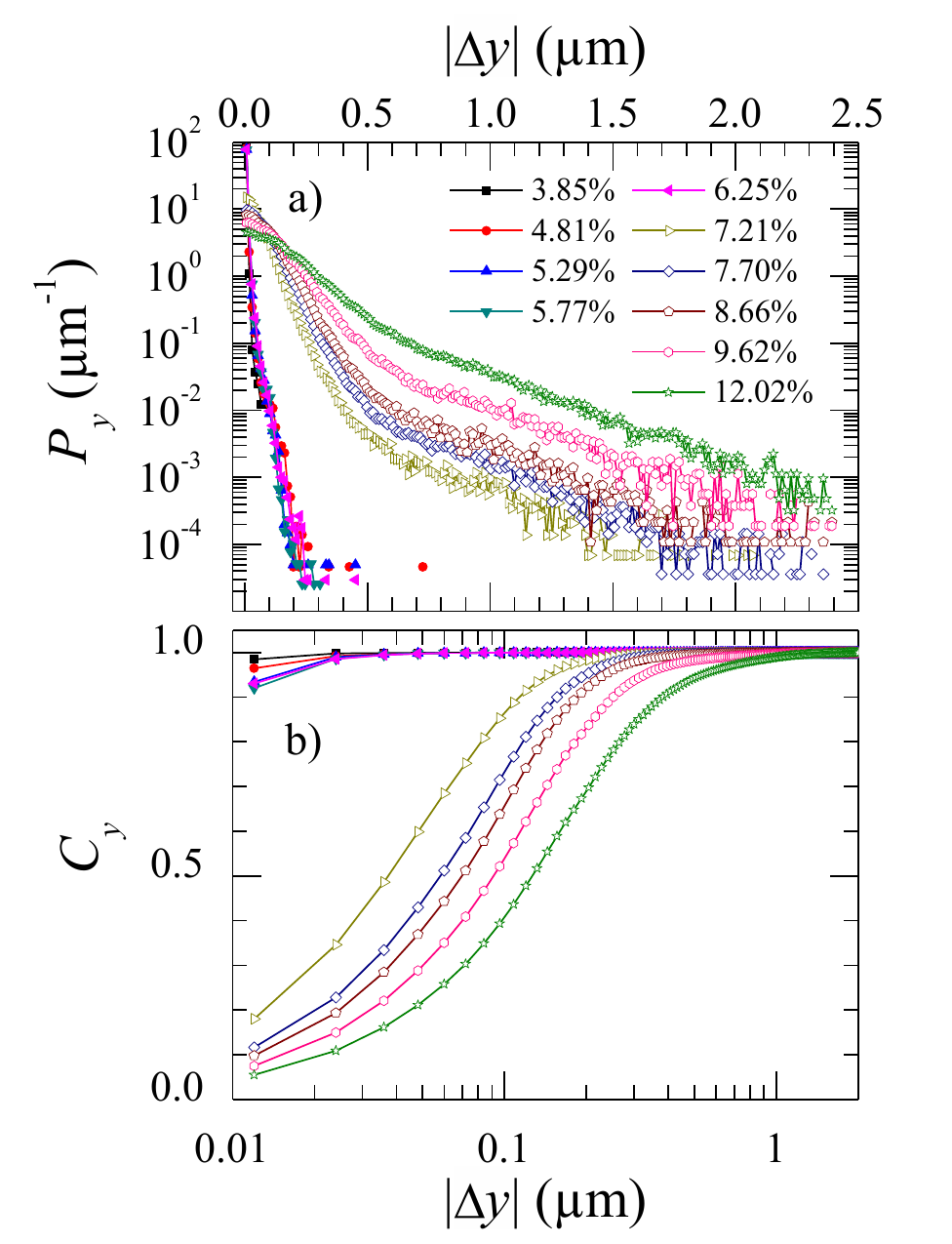}
  \caption{(a) Probability distribution function of the drop displacements over one shear cycle, in the direction orthogonal to the applied strain, for an emulsion at $\varphi = 0.74$. Curves are labeled by the amplitude of the applied strain. Solid (resp., open) symbols refer to $\gamma$ below (resp., above) the yield strain $\gamma_c$. (b) Cumulative distributions obtained by integrating the probability distributions shown in the top panel.}
  \label{fig:74stat}
\end{figure}
%------------------------------------------------------------------ FIG

To investigate quantitatively the heterogeneity of the dynamics hinted to by the images of Fig.~\ref{fig:differences} , we calculate the probability distribution function of the drop displacement over one cycle, $P_y(\Delta y)$. Figure~\ref{fig:74stat}(a) shows $P_y$ for a sample at $\varphi=0.74$: its features are representative of all samples at $\varphi \ge 0.74$, for which the yielding transition is very sharp. We find that positive and negative displacements are equally likely; we therefore plot the data against $|\Delta y|$. Two well-separated sets of curves are visible. For strains smaller than the yield strain (solid symbols), the pdf's decay very rapidly to zero, implying that most of the drops are essentially quiescent. This is also shown in Fig.~\ref{fig:74stat}(b), where we plot the cumulative distribution function, $C_y(|\Delta y|) = \int_0^{|\Delta y|}P_y(u)\mathrm{d}u$. For $\gamma < \gamma_\mathrm{y} = 6.32\%$, 98.5\% or more of the drops move by less than $0.024\um$, \textit{i.e.} a mere one-hundreth of their size. By contrast, for $\gamma > \gamma_y$ (open symbols), motion on a much larger length scale is observed. As seen in Fig.~\ref{fig:74stat}(a), above the yield strain the pdf's are characterized by a two-step shape, with two distinct decays that appear as straight lines in a semilog plot. Thus, two populations of particles with different mobility coexist, which we shall refer to as the `mobile' and `supermobile' particles, each population being characterized by a nearly exponential distribution of displaments. It should be noted that a non-negligible fraction of particles still remain almost quiescent. Indeed, the cumulative distributions of Fig.~\ref{fig:74stat}(b) reveal that at $\gamma = 7.21\%$, just above the yielding transition, more than a third of the particles (34.6\%) move less than $0.024\um$, the restrained displacement typical of the vast majority of drops below $\gamma_\mathrm{y}$. Even at the largest applied strain, $\gamma = 12.02\%$, more than 10\% of the particles remain essentially quiescent. The persistence of a fraction of immobile droplets is consistent with the quiescent patches observed, at all strains, when subtracting pairs of consecutive images, as in Fig.~\ref{fig:differences}. It is also consistent with previous experiments by both DWS~\cite{hebraud1997} and microscopy.\cite{Clara-Rahola2012}

Figure~\ref{fig:74stat}(a) reveals the origin of the continuous growth of the rms displacement beyond the yielding transition seen in Fig.~\ref{fig:transition}(a). As $\gamma$ increases, two concomitant mechanisms lead to a larger average displacement. On the one hand, the typical displacement of the mobile particles grows, as indicated by the flattening of the first decay in the pdf's. On the other hand, the fraction of supermobile particles increases, while their characteristic displacement remains roughly constant, as indicated by the fact that the large-$|\Delta y|$ tails of the distributions have nearly the same slope but are shifted upwards with increasing $\gamma$. By fitting the tail of the pdf's to an exponential decay, $P_y \sim \exp(-|\Delta y|/\Delta_\mathrm{s})$, we find the characteristic displacement of the supermobile particles to be $\Delta_\mathrm{s} = 0.31\um$, where the quoted value has been obtained by averaging the results for $\gamma \ge 7.7\%$,  which exhibit little variation with $\gamma$. Such a displacement corresponds to about 13\% of the average drop size.

%------------------------------------------------------------------- FIG
\begin{figure}[h]
\centering
  \includegraphics[width=1\columnwidth,clip]{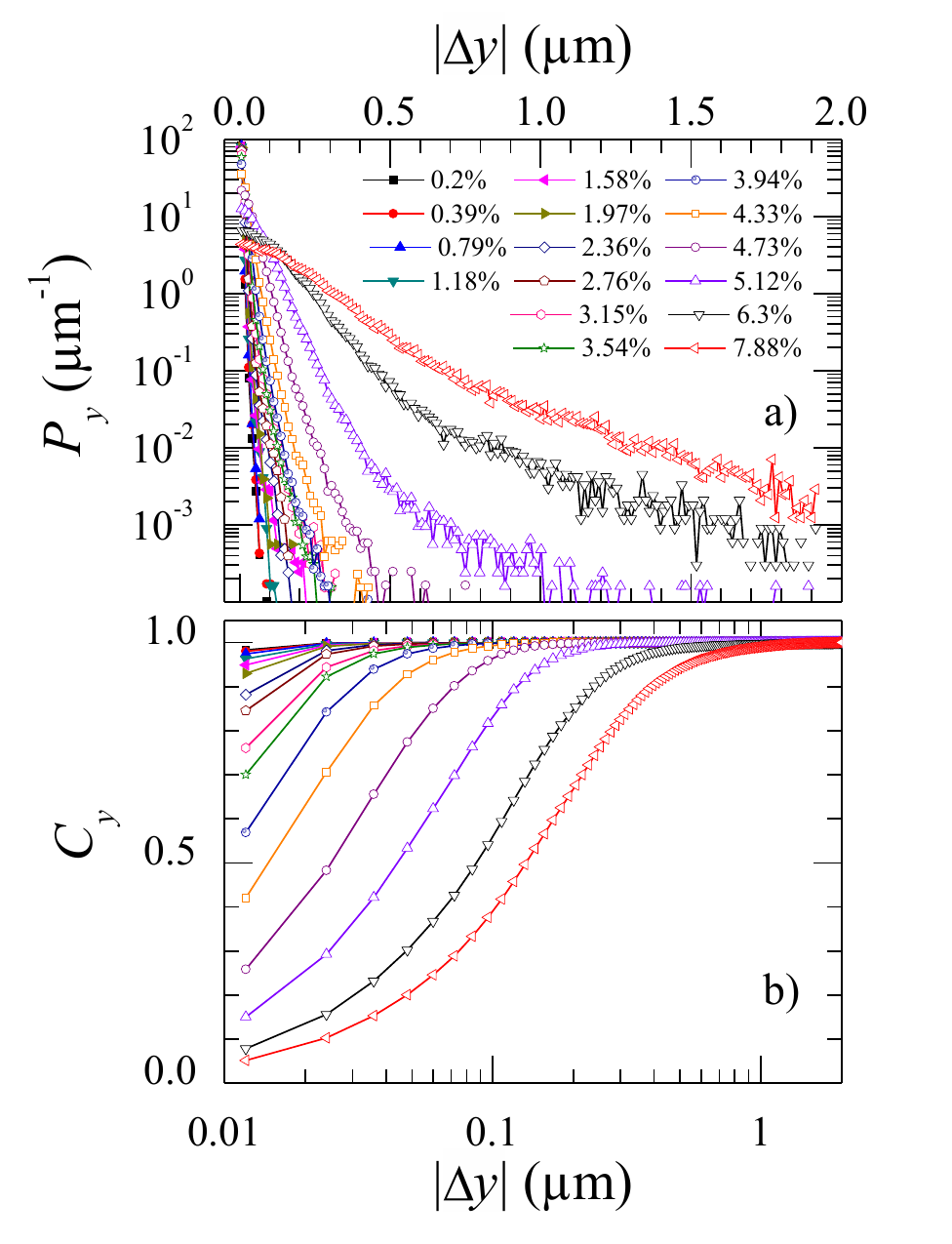}
  \caption{(a) Probability distribution function of the drop displacements over one shear cycle, in the direction orthogonal to the applied strain, for an emulsion at $\varphi = 0.65$. Curves are labeled by the amplitude of the applied strain. Solid (resp., open) symbols refer to $\gamma$ below (resp., above) the yield strain $\gamma_c$. (b) Cumulative distributions obtained by integrating the probability distributions shown in the top panel.}
  \label{fig:65stat}
\end{figure}
%------------------------------------------------------------------ FIG

Figure~\ref{fig:65stat} shows the probability (a) and cumulative (b) distributions of the drop displacements for the most diluted sample ($\varphi = 0.65$), for which the least sharp yielding transition was observed in Fig.~\ref{fig:transition}. Consistent with a gentler transition, we find that both $P_y$ and $C_y$ evolve in a more continuous way with $\gamma$, as compared to the sample at $\varphi = 0.74$ shown in Fig.~\ref{fig:74stat}: no clear separation in the two family of curves is observed. For the sample at $\varphi = 0.70$, the behavior is intermediate between those shown in Figs.~\ref{fig:74stat} and~\ref{fig:65stat} (data not shown). In spite of the differences for $\gamma \le \gamma_\mathrm{y}$, the behavior beyond the yielding transition is very close for all samples, including the most dilute one shown in Fig.~\ref{fig:65stat}. Again we find that quiescent particles coexist with two distinct populations of mobile and supermobile droplets. Similarly to the case of the sample at $\varphi = 0.75$, the typical displacement of the supermobile particles is insensitive to the amplitude of the applied strain, while the fraction of supermobile particles increases with $\gamma$, until the pdf exhibits almost a single exponential decay, for the largest $\gamma$. From an exponential fit to the tail of the pdfs' at large $\gamma$, we obtain $\Delta_\mathrm{s} = 0.35\um$, comparable to the typical displacement of the supermobile particles measured at $\varphi = 0.74$.

%------------------------------------------------------------------- FIG
\begin{figure}[h]
\centering
  \includegraphics[width=1\columnwidth,clip]{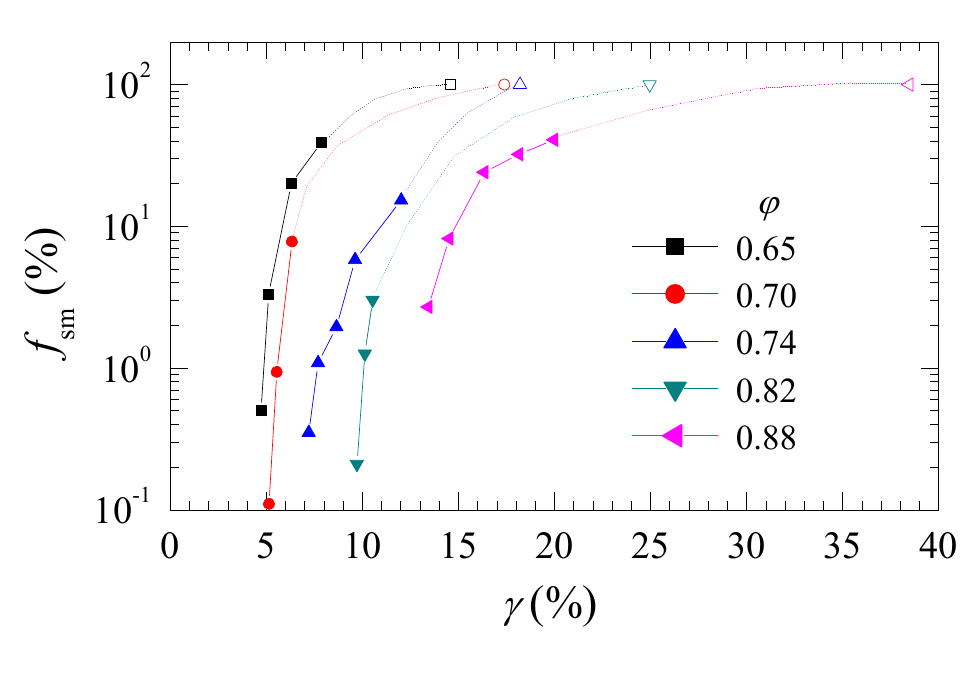}
  \caption{Solid symbols: fraction $f_\mathrm{sm}$ of the ``supermobile'' drops, whose displacement falls in the the exponential tail of the probability distribution $P_y$, as discussed in the text. The abscissa of the open points correspond to the rheological crossover strain $\gamma_\mathrm{f}$, where the sample is fluidized. The dotted lines are guides to the eye, consistent with the hypothesis that full fluidization may occur when all drops are supermobile, \textit{i.e.} for $f_\mathrm{sm}=100\%$.}
  \label{fig:supermobilefraction}
\end{figure}
%------------------------------------------------------------------ FIG

From a similar analysis performed on the pdfs for all investigated $\varphi$, we find that the typical displacement of the supermobile particles is $\overline{\Delta_\mathrm{s}} = (0.26\pm0.07 )\um$, where the average and the uncertainty are calculated over all experiments at large strain, regardless of $\varphi$. Quite intriguingly, the average $\Delta_\mathrm{s}$ is about 11\% of the drop size, close to the amplitude of particle motion leading to the melting of a crystalline solid according to Lindemann's criterion~\cite{lindemann1910}, for which the typical threshold is around 15\% of the particle spacing. Although the supermobile particles are just a small fraction of all drops at the yielding transition, their relative number increases rapidly as $\gamma$ exceeds $\gamma_\mathrm{y}$ (see the raise of the tails of $P_\mathrm{y}$ in Figs.~\ref{fig:74stat}(a) and~\ref{fig:65stat}(a)). The question naturally arises as whether the full fluidization of a sheared emulsions occurs when all (or at least the majority of) the particles are supermobile, which would constitute an analogous of the Lindemann's criterion for an amorphous solid.
To test this idea, we plot as solid symbols in Fig.~\ref{fig:supermobilefraction} the fraction of supermobile particles, as a function of the amplitude of the applied strain, where a particle is taken to be supermobile if its displacement over one shear cycle exceeds $\overline{\Delta_\mathrm{s}}$. We add to the plot the points that would correspond to a full fluidization when all particles are supermobile (open symbols): by definition, their ordinate is 100\%, while their abscissa is $\gamma_\mathrm{f}$, the fluidization strain beyond which $G'' > G'$. The dotted lines are a guide to the eye, connecting the solid and open symbols. While our experiments do not allow this extension of Lindemann's criterion to be definitively confirmed, the data are sufficiently compelling to call for more experiments in a wider range of applied strains.

%------------------------------------------------------------------- FIG
\begin{figure}[h]
\centering
  \includegraphics[width=1\columnwidth,clip]{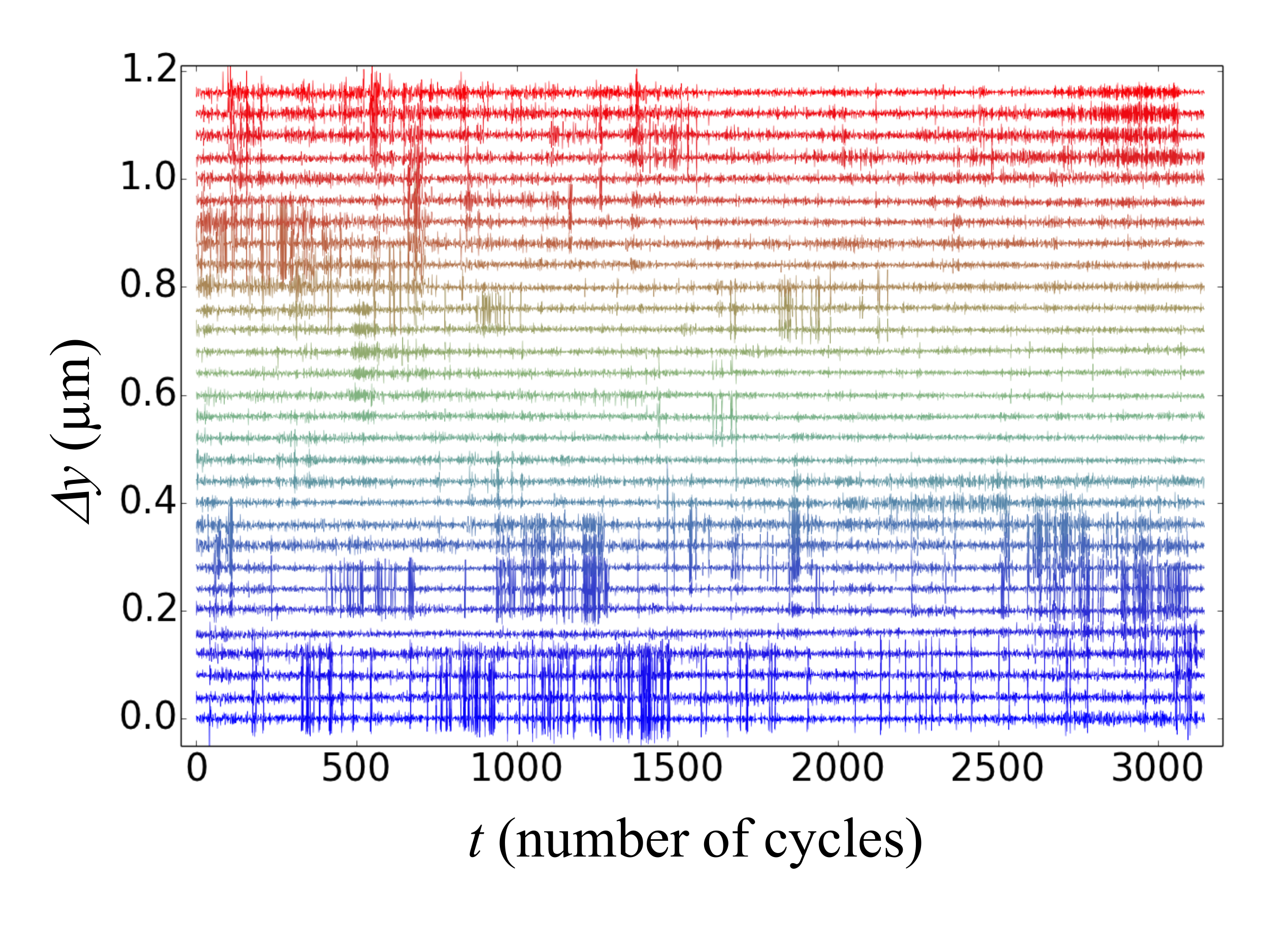}
  \caption{Temporal evolution of $\Delta y$, measured for 30 ROIs aligned along a vertical line spanning the whole height of the microscopy images. Data for $\varphi = 0.7$ and $\gamma = 6.25\%$, just below the yielding transition. Each curve has been offset by $0.04\um$ with respect to the previous one for visibility.}
  \label{fig:timeseries}
\end{figure}
%------------------------------------------------------------------ FIG

We now turn to the spatio-temporal organization of the droplet motion. Figure~\ref{fig:timeseries} shows a set of time series of the drop displacement, $\Delta y(t,\mathrm{\mathbf{R}})$, measured in 30 ROIs whose center $\mathrm{\mathbf{R}}$ is aligned along a vertical line spanning the whole height of the microscopy images, for a sample at $\varphi = 0.7$ and for $\gamma = 6.25\%$, just below the yielding transition. Two features strike the eyes: on the one hand droplet motion is not continuous in time, but rather proceeds by intermittent bursts of activity followed by quiescent periods. On the other hand, regions with correlated dynamical activity extend over several ROIs, since the time series for nearby ROIs exhibit a similar behavior. We go beyond this qualitative observations by measuring both the temporal and spatial correlations of the dynamics.

%------------------------------------------------------------------- FIG
\begin{figure}[h]
\centering
  \includegraphics[width=1\columnwidth,clip]{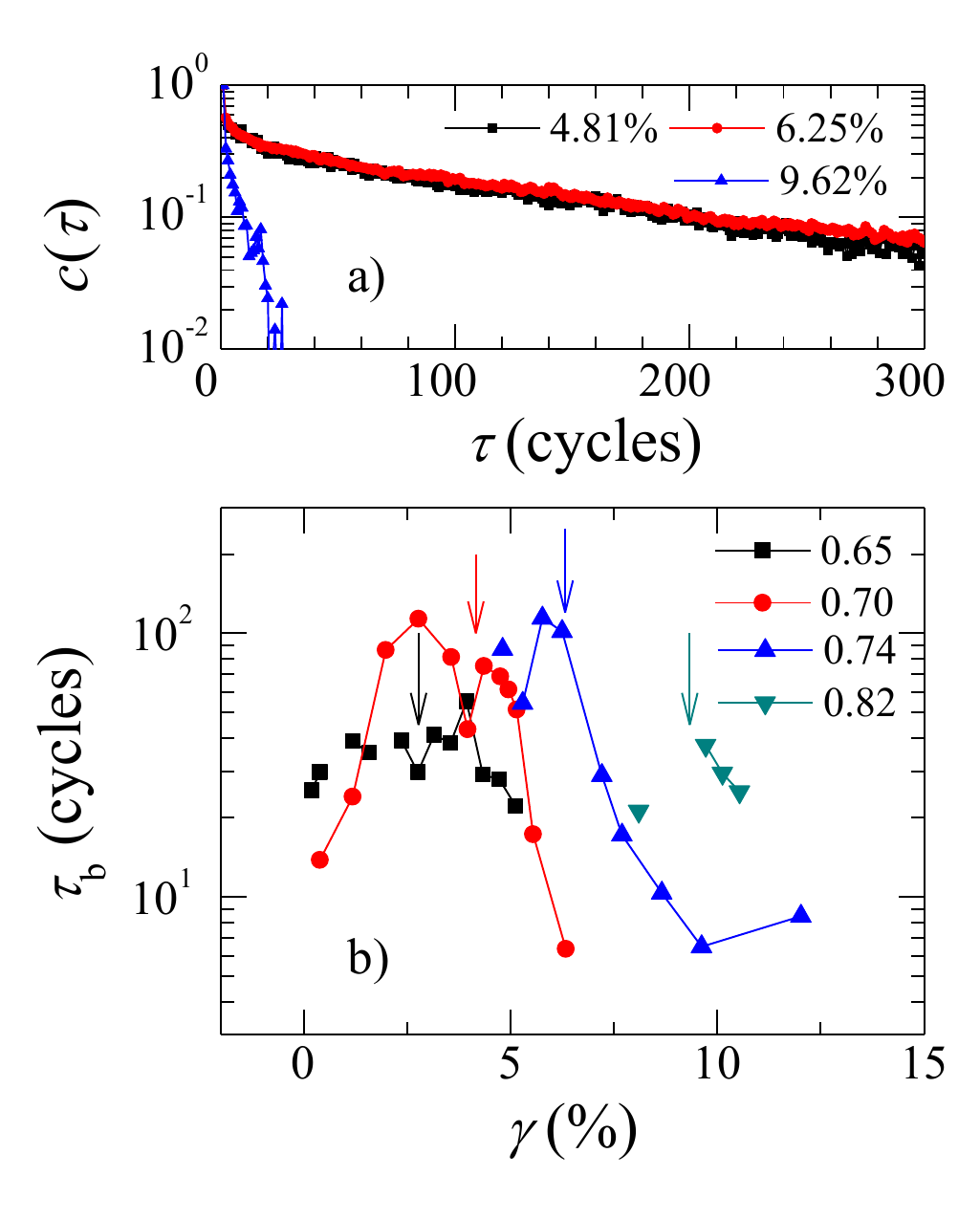}
  \caption{(a) Time autocorrelation function of $|\Delta y(t)|$, for a sample at $\varphi = 0.74$ and for the three strain amplitudes shown in the label. $\gamma$ is well below, just below and well above the microscopic yield strain $\gamma_\mathrm{c}$, respectively. (b) Strain dependence of the characteristic time of $c(\tau)$, for samples at $\varphi \le 0.82$, as indicated by the label. The autocorrelation functions for $\varphi = 0.88$ are too noisy for a relaxation time to be extracted reliably. The arrows indicate the location of the microscopic yielding transition.}
  \label{fig:timecorr}
\end{figure}
%------------------------------------------------------------------ FIG

Temporal correlations are quantified by calculating
\begin{equation}
\widetilde{c}(\tau) = \left < \left <|\Delta y(t,\mathrm{\mathbf{R}})||\Delta y(t+\tau,\mathrm{\mathbf{R}})| \right>_t \right >_\mathbf{R}
\end{equation}
where $ \left < \cdot \cdot \cdot \right >_t$ and $ \left < \cdot \cdot \cdot \right >_{\mathbf{R}}$ indicate averages over the experiment duration and the ROI location, respectively. Note that we calculate the autcorrelation of the absolute $y$ displacement, rather than that of $\Delta y$ itself. This is because we aim at capturing the duration of the activity bursts, better seen when the rapid and chaotic fluctuations around 0 of the displacement are suppressed. We find that $\widetilde{c}(\tau)$ still contains significant noise contributions at time lags $\tau = 0$ and $\tau = 1$. We thus neglect the first two points of $\widetilde{c}(\tau)$ and study the normalized time autocorrelation function, defined as $c(\tau) = \widetilde{c}(\tau)/\widetilde{c}(2)$, for $\tau \ge 2$. Typical examples of $c$ are shown in Fig.~\ref{fig:timecorr}(a), for $\varphi = 0.74$ and three strain amplitudes, well below, near to, and well above the yielding transition, respectively. We extract the characteristic decay time of the autocorrelation function, which we identify with the typical burst duration $\tau_\mathrm{b}$, from
\begin{equation}
\tau_\mathrm{b} = \frac{1}{T-2} \int_2^T c(\tau)\mathrm{d}\tau  \,,
\end{equation}
where $T$ is the smallest delay for which $c$ has decayed below 0.1. The strain dependence of $\tau_\mathrm{b}$ is shown in Fig.~\ref{fig:timecorr}(b), for various $\varphi$. For the largest volume fraction, $\varphi = 0.88$, $\widetilde{c}$ decays very rapidly to a noisy base line, thus making our normalization procedure difficult to apply and preventing $\tau_\mathrm{b}$ for being calculated. Overall, $\tau_\mathrm{b}$ is quite large, from several cycles to more than 100 cycles: this highlights  the need of performing long experiments in order to fully capture the temporal behavior of the rearrangements. On the other hand, the burst duration remains finite: this rules out the hypothesis that a given fraction of the drops are \textit{always} immobile, while some of them are \textit{always} rearranged, as proposed in the DWS experiments of~\citenst{hebraud1997}. It should be noted however that in Ref.~\citenst{hebraud1997} the average dynamics could only be measured over a period of 16 cycles: in most cases such a reduced time span is comparable to or shorter than $\tau_\mathrm{b}$, which would explain why the populations of mobile and immobile particles did not seem to be exchanged over time in the DWS experiments. %This is precisely what I suspected, so this is a very helpful result! (DJP)

%Although for $0.70 \le \varphi \le 0.82$ the temporal correlation of the dynamics appears to be maximum around $\gamma_\mathrm{y}$, no explosive growth is seen around the transition, as one may have expected if the yielding transition was second-order.

Spatial correlations of the dynamics are investigated by inspecting how similar are the displacement time series recorded
in two locations of the sample, as a function of the location separation $\Delta \mathbf{r}$.
%We have checked that similar results are obtained when analyzing the components of the drops' displacement parallel or perpendicular to the shear flow.
We % thus focus on the spatial correlation of the displacements along $y$ and
introduce the un-normalized spatial correlation $G_{4}(\Delta \mathbf{r})$, similar to those used to analyze heterogeneous dynamics in glass formers and granular media~\cite{dhbook}, and defined by
%+++++++++++++++++++++++++++++++++++++++++++
\begin{equation}
G_{4}(\Delta \mathbf{r}) = \left < \left < \Delta y(\mathbf{R},t)\Delta y(\mathbf{R}+\Delta \mathbf{r},t)\right >_t \right >_\mathbf{R}  \,,
\label{eq:G4}
\end{equation}
%+++++++++++++++++++++++++++++++++++++++++++++++
%where $\Delta y(\mathbf{r},t)$ is the $y$ component of the displacement between the images taken at time $t$ and $t+T$, measured for a ROI located in $\mathbf{r}$, and where $ \left < \cdot \cdot \cdot \right >_t$ and $ \left < \cdot \cdot \cdot \right >_{\mathbf{r}}$ indicate averages over the experiment duration and the ROI location, respectively.
Note that $G_4$ compares not only the magnitude of the displacement, but also the \textit{direction} along which it occurs, since the correlation is
calculated for the $y$ component, rather than for the  modulus of the displacement. Thus, $G_4$ quantifies spatial correlations in a rather
stringent sense, rather than just dynamical activity.
We find that  $G_{4}$ does not depend on the orientation of $\Delta \mathbf{r}$, \textit{i.e.} that spatial correlations are isotropic, consistent with the observation that the plastic dynamics observed in our experiments are independent of the shear orientation. In the following we thus report the azimuthally-averaged $G_4$, which we normalize by calculating
%+++++++++++++++++++++++++++++++++++++++++++
\begin{equation}
g_{4}(\Delta r) = a\left [G_4(\Delta r) - b\right ] \,.
\label{eq:g4}
\end{equation}
%+++++++++++++++++++++++++++++++++++++++++++++++
The coefficients $a$ and $b$ are chosen such that $g_4(\Delta r \rightarrow 0)=1$, $g_4(\Delta r \rightarrow \infty)=0$. In practice, $b$ is obtained as the mean value of $G_4$ for $\Delta r \ge 96 \um$, while $a$ is obtained as the 0-th order term in a cumulant fit~\cite{koppel1972} to $G_4 - b$:
\begin{equation}
\ln[G_4(\Delta r)-b] = \ln a - \xi_y \Delta r + \Gamma_2 (\Delta r)^2 \,.
\label{eq:cumulants}
\end{equation}

%------------------------------------------------------------------- FIG
\begin{figure}[h]
\centering
  \includegraphics[width=0.95\columnwidth,clip]{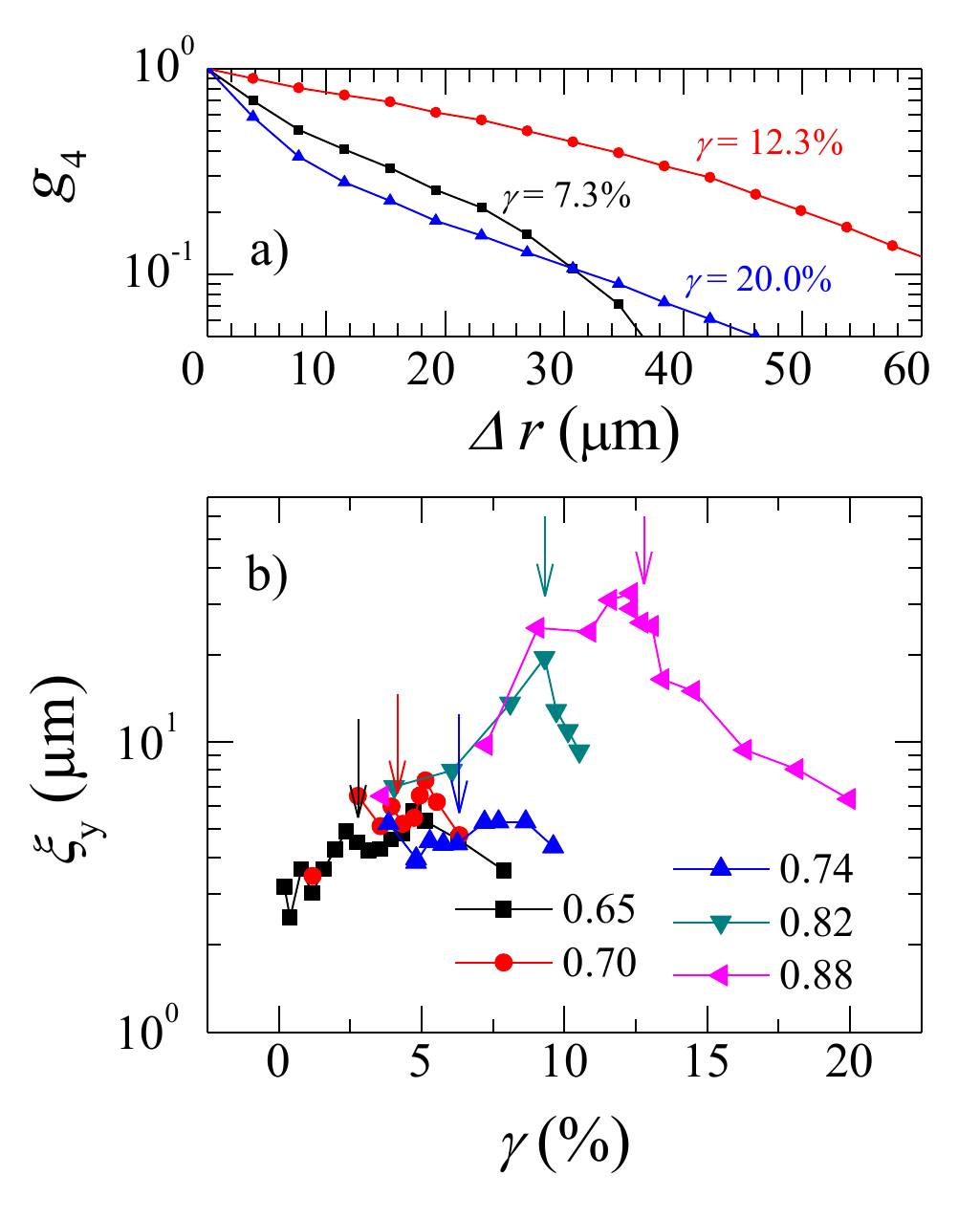}
  \caption{(a) Spatial correlation of the dynamics ($y$ component), for a sample at $\varphi = 0.88$ and for the three strain amplitudes shown by labels. $\gamma$ is well below, nearly at and well above the microscopic yield strain $\gamma_\mathrm{c}$, respectively. (b) Strain dependence of the correlation length of the dynamics, obtained from fits to $g_4$. Data are labeled by $\varphi$. The arrows indicate the location of the microscopic yielding transition.
  }
  \label{fig:spacecorr}
\end{figure}
%------------------------------------------------------------------ FIG

Figure~\ref{fig:spacecorr}(a) shows $g_4$ for the most concentrated sample ($\varphi = 0.88$), for three strain amplitudes well below, nearly at, and well above the yielding transition, respectively. At both small and large applied strain, the dynamics are correlated over a few microns, a distance comparable to a couple of drop sizes. By contrast, much long range spatial correlations of the dynamics are observed close to the microscopic yield strain. To test the generality of this behavior, we show in Fig.~\ref{fig:spacecorr}(b) for all samples the $\gamma$ dependence of $\xi_y$, the characteristic range of the spatial correlations obtained from cumulant fits, Eq.~(\ref{eq:cumulants}). For $\varphi \le 0.74$, the behavior of $\xi_y$ is very different from that inferred from Fig.~\ref{fig:spacecorr}(a): at the lowest volume fractions, the range of correlated motion remains modest and quite insensitive to $\gamma$ over the full range of applied strain amplitudes. At larger $\varphi$, by contrast, $\xi_y$ grows significantly as the applied strain approaches the yield value, where spatial correlations may extend up to 20 particle sizes. At even larger strain, $\xi_y$ decreases. Inspection of movies of the strobed images reveal the origin of this non-monotonic behavior. As $\gamma$ is increased, coordinated motion of increasingly large patches of particles is observed, until essentially all the sample moves, but in an increasingly disordered way at very high $\gamma$. %As for the temporal correlations of the dynamics, we note that no explosive growth of $\xi$ is observed around the yielding transition, again in contrast with the notion of a second-order dynamical transition to irreversibility.

\section{Discussion and conclusions}
\label{sec:conclusions}

The first important finding of our experiments is the steep increase of the drop mobility over a restricted range of applied strains, shown in Fig.~\ref{fig:transition}(a), which contrasts with the very smooth yielding transition observed by rheology (Fig.~\ref{fig:rheo}). This behavior likely depends on the fact that, in a compressed emulsion, local motion of a few drops is inevitably propagated elastically to a large number of neighbors. This would favor a nearly on-off behavior, where either the sample is (almost) quiescent everywhere, or it is significantly rearranged (almost) everywhere. Two observations support this view. First, as $\varphi$ decreases, the transition become less sharp, presumably because this mechanism is less efficient just above random close packing, where elastic propagation of a local rearrangement is screened. Second, a similar trend was reported for sheared colloidal glasses.\cite{PetekidisPRE2002} It should be noted that the colloidal glasses of Ref.~\citenst{PetekidisPRE2002} were prepared at lower volume fractions than our emulsions. Consistent with the proposed scenario, the yielding transition is found to be overall smoother for colloidal glasses than in the experiments reported here. In spite of the differences in the sharpness of the microscopic and rheological yielding transitions, the two critical strains, $\gamma_\mathrm{y,r}$ and $\gamma_\mathrm{y}$ obtained from the two techniques agree remarkably well. This result appears to be robust with respect to the criterion used to define the microscopic yield strain, since a similar conclusion was also reached by analyzing the DWS intensity correlation functions in Ref.~\citenst{hebraud1997}. It should be emphasized that $\gamma_\mathrm{y}$ is much smaller than the fluidization strain $\gamma_\mathrm{f}$: substantial plastic rearrangements occur well before the sample flows.

Another important finding is the fact that the dynamics are heterogenous, with  almost quiescent particles coexisting with mobile particles. Indeed, as discussed in reference to Figs.~\ref{fig:74stat} and ~\ref{fig:65stat}, a significant fraction of drops undergo very restricted motion even around or above the yield strain. This is consistent with the interpretation of the DWS measurements proposed in Ref.~\citenst{hebraud1997}. However, the microscopy experiments allow a more refined picture of the drop dynamics to be drawn. A very peculiar distribution of the amplitude of the drop motion is seen, as evidenced by the exponential tails of $P_y$ in Figs.~\ref{fig:74stat}(b) and ~\ref{fig:65stat}(b). We recall that for diffusive motion $P_y$ should be Gaussian, a behavior clearly incompatible with our data. Exponential tails in the pdf's of the displacement have been consistently observed in glassy and jammed systems, \textit{e.g.} in sheared foams,\cite{mobius2010} grains,\cite{MartyPRL2005} and colloids,~\cite{ChenPRE2010} and even in simulations of molecular glass formers.\cite{chaudhuri2007} A general mechanism has been proposed to explain these exponential tails,\cite{chaudhuri2007} based on the idea that particles undergo discontinuous jumps. This picture seems compatible with the nature of the rearrangements in our emulsions. However, the double exponential tail measured here points to a more complex behavior, which will need further experiments to be fully understood, for example by studying systematically the spatial and temporal distribution of the supermobile particles.

The droplet dynamics show complex spatio-temporal organization. While such heterogenous behavior is overall consistent with what inferred in Ref.~\citenst{hebraud1997},
an important difference concerns the life time of the active regions, which in the optical microscope experiments is found to be finite, albeit very long: $\tau_\mathrm{b}$
can be as long as hundreds of cycles, which highlights the necessity of performing experiments over a very large number of cycles.
Both the time scale and the spatial range of dynamical correlations have a non-monotonic behavior with $\gamma$, reaching
a maximum at a strain comparable to the yield strain, except for the lowest $\varphi$, for which little variation of $\tau_\mathrm{b}$
and $\xi_y$ is observed when varying $\gamma$. In recent simulations of a sheared Lennard-Jones glass,\cite{Priezjev2013} a monotonic
increase of the range of spatial correlations with $\gamma$ was reported. Unfortunately, the largest tested strain, $\gamma = 6\%$,
was relatively limited, so that it is difficult to conclude as whether or not a non-monotonic behavior would be observed at even larger strains.
Non-monotonic spatial correlations as a function of $\varphi$ have been also reported for the spontaneous dynamics of colloids close
to jamming~\cite{BallestaNatPhys2008}. Close to random close packing, spatial correlations of the dynamics extend over a few particles
at most, similarly to Ref.~\citenst{Priezjev2013} and in analogy to the dynamics of other glassy or jammed systems, both driven and at rest.\cite{dhbook}
As $\varphi$ grows, the range of spatial correlations significantly grows, as seen in Fig.~\ref{fig:spacecorr}(b).
The same trend, with similar values of $\xi$ (expressed in numbers of drops) was observed in Ref.~\citenst{goyon2008}.
This analogy is remarkable, given that the experiments of Goyon \textit{et al.} are performed in a steady flow regime,
where in our case the largest $\xi_y$ are measured around $\gamma_\mathrm{y}$, when the sample has still a predominantly solid-like behavior.

Finally, it is interesting to consider the experiments presented here from the perspective of the ongoing debate on the transition from reversibility to
irreversibility in driven complex fluids.\cite{pine2005,corte2008,SlotterbackPRE2012,keim2013,SastryPRE2013,jeanneret2014} A first point to keep in
mind is that no fully reversible state is actually observed here, since even at the lowest applied strains some dynamics are observed. This is also
true, at least on a finite time scale, for most of the work on systems of interacting particles where the existence of a reversible regime was
advocated.\cite{SlotterbackPRE2012,keim2013,SastryPRE2013,jeanneret2014} This has to be contrasted with the experiments on dilute,
non-interacting particles,\cite{pine2005,corte2008} where a genuine fully reversible state was observed. Concerning the nature of the
transition, for our system we find contrasting results according to $\varphi$. Close to random close packing, the smooth increase of
the average drop displacement and the continuous evolution of $P_y$ with $\gamma$
are suggestive of a second order-transition, but no strong growth of temporal and spatial quantities is seen around the transition, in contrast to
the findings of Refs.~\citenst{pine2005,corte2008}. At higher volume fraction, both  $\tau_\mathrm{b}$ and $\xi_y$ grow significantly when approaching the
 yielding transition, but the partitioning between mobile and (almost) immobile particles and the sharpness of the transition are rather suggestive
 of a first-order transition, as in Ref.~\citenst{jeanneret2014}. Overall, it seems that the nature of the yielding transition depends on the distance
 from the jamming transition and that the analogy with the transition to irreversibility discussed for diluted suspensions~\cite{pine2005,corte2008}
  may be of limited relevance.

Several questions are left open by the work presented here and will deserve further investigation. The existence of a Lindemann's criterion for amorphous solids
discussed in relation to Fig.~\ref{fig:supermobilefraction} is appealing, but certainly needs additional experiments at larger $\gamma$ to
be fully assessed. A second research direction would be exploring the dependence of the dynamics on time delay, $\tau$. While exploring motion after
one full shear cycle, as discussed here, is an essential starting point, particularly relevant for the comparison with rheology data, it is clear
that much is to be learned by analyzing how particle displacement grows with time. A related question concerns the length scale dependence
of the dynamics. While numerical works generally suggest that plastic rearrangements lead to diffusive dynamics, very recent experiments on a
sheared colloidal polycrystal have shown that the network of defects relaxes \textit{ballistically} under an oscillatory shear.\cite{tamborini2013} It would
be interesting to investigate the nature of the dynamics in concentrated emulsions, for which ballistic dynamics have been reported in the past for
a sample submitted to compressional stress in a centrifuge.\cite{LucaFaraday2003}

\section*{Acknowledgments}

This work was supported by U.S. National Science Foundation (Grant No. CTS-0221809) and CNES. LC thanks NYU Center for Soft Matter Research for supporting his stay at NYU. We acknowledge fruitful discussions with L. Berthier.

\providecommand*{\mcitethebibliography}{\thebibliography}
\csname @ifundefined\endcsname{endmcitethebibliography}
{\let\endmcitethebibliography\endthebibliography}{}

%%\footnotesize{
%%\bibliography{drops}
%your .bib file
%%\bibliographystyle{rsc}
%the RSC's .bst file
%%}

\end{document}